\RequirePackage{fix-cm}
\documentclass[twocolumn]{svjour3}          
\smartqed  
\usepackage{float}
\usepackage{amsfonts}
\usepackage{subfigure}
\usepackage{tikz}
\usepackage{graphicx}
\usepackage{mathtools, cuted}
\usepackage{microtype}
\usepackage{csquotes}
\usepackage{adjustbox}
\usepackage{dirtytalk}
\usepackage{ctable}
\usepackage{booktabs}
\usepackage{longtable}
\usepackage{paralist, tabularx}
\DeclareMathOperator*{\argminA}{arg\,min} 
\begin{document}
	
	\title{Performance-based Post-earthquake Decision-making for Instrumented Buildings
	}
	
	
	\author{Milad Roohi  \and Eric M. Hernandez }
	
	
	\institute{\and NIST Center for Risk-based Community Resilience Planning, Department of Civil and Environmental Engineering, Colorado State University, Fort Collins, CO 80521 USA. \email{mroohigh@colostate.edu} \\ 
		\and
		Department of Civil and Environmental Engineering, University of Vermont, Burlington, VT 05405 USA.
		\email{eric.hernandez@uvm.edu} }
	
	\date{Received: date / Accepted: date}
	
	\maketitle
	
	\begin{abstract}
		This paper presents a framework for decision-making regarding post-earthquake assessment of instrumented buildings in a manner consistent with performance-based design criteria. This framework is achieved by simultaneously combining and advancing existing knowledge from seismic structural health monitoring and performance-based earthquake engineering paradigms and consists of 1) optimal sensor placement, 2) dynamic response reconstruction, 3) damage estimation, and 4) performance-assessment and decision-making. In particular, the main objective is to reconstruct inter-story drifts with a probabilistic measure of exceeding performance-based acceptance limits and determining the post-earthquake re-occupancy classification of the instrumented building of interest.  Since the proposed framework is probabilistic, the outcome can be used to obtain the probability of losses based on the defined decision variables and be integrated into a risk-based decision-making process by city officials, building owners, and emergency managers. The framework is illustrated using data from the Van Nuys hotel testbed, a seven story reinforced concrete building instrumented by the California Strong Motion Instrumentation Program (Station 24386). 
		
	\end{abstract}
	
	\section{Introduction}
	\label{intro}
	
	After a potentially damaging earthquake, city officials must make decisions regarding the structural integrity of building structures under their jurisdiction. Furthermore, exposure to sequential seismic events following a major earthquake can exacerbate the problem resulting in cumulative physical damage and some cases even partial or complete collapse. Recent earthquakes in New Zealand (2010–2011 Canterbury sequence), Taiwan (1999 Chi-Chi mainshock-large aftershocks sequence), Iran (2012 East Azerbaijan doublet) have demonstrated that this is not a hypothetical scenario, but a real possibility. Although most engineered buildings are expected to survive a strong ground motion without collapse, it is not clear that these buildings will be safe to re-occupy, especially if they can be subjected to strong aftershock. 
	
	In the United States, and other parts of the world, documents such as ATC-20 \cite{ATC20_1989} and ATC-20-2 \cite{ATC20_1995} (by Applied Technology Council) offer guidelines for visual post-earthquake assessment and occupancy classification of potentially damaged buildings as inspected (green tag), restricted use (yellow tag) and unsafe (red tag). Despite best efforts by inspectors, visual inspections suffer from several notable limitation, including but not limited to: (1) inspector bias and (or) experience-based variability, (2) lack of access to damaged locations or members, (3) time consuming, and (4) qualitative in nature and not entirely quantitative or physics-based. These limitations might lead inspectors to reach erroneous conclusions about which buildings are safe to be re-occupied immediately and which ones are not, exacerbating earthquake losses. 
	
	In the case of instrumented buildings, engineers can augment the assessment by incorporating measurements during earthquakes. However, despite the immediate appeal, there are technical, logistical, and economic challenges associated with building instrumentation including but not limited to (a) it is not possible to measure damage directly during and following an earthquake, (b) building instrumentation and its maintenance are relatively expensive, and budget constraints may not allow floor-by-floor or component-level instrumentation of a building, and (c) in practice, due to budget constraints, only a limited number of accelerometers (an average of 9 to 12) are installed per building. Therefore, it is required to develop a rapid and reliable decision-making procedure to integrate all the available information (such as measurements, structural drawings, construction information) to assess the extent of the structural damage indirectly and, subsequently, make informed decisions. Such a procedure will lead to mitigate earthquake losses, reduce the decision-making uncertainty, and improve community resilience.
	
	This paper address this need by presenting a framework for decision-making regarding the post-earthquake assessment of instrumented buildings in a manner consistent with criteria from performance-based design. This framework is achieved by simultaneously combining and advancing existing knowledge from seismic structural health monitoring and performance-based earthquake engineering paradigms and consists of 1) optimal sensor placement, 2) dynamic response reconstruction, 3) damage estimation, and 4) performance-assessment and decision-making. Since the proposed framework is developed on a probabilistic basis, the outcome can be used to obtain the probability of various losses based on the defined decision variable and be integrated into a risk-based decision-making process by city officials, building owners, emergency managers, or other officials. 
	
	This paper is organized as follows. Section \ref{Section2} presents the system and measurement models of interest. This is followed by section \ref{Section3} that presents background on performance-based earthquake engineering. Section \ref{Section4} develops the proposed framework for performance-based post-earthquake decision making. Finally, the paper ends with a case study of the Van Nuys hotel testbed, a seven story reinforced concrete building instrumented by the California Strong Motion Instrumentation Program (Station 24386), to illustrate the effectiveness of the proposed framework.
	
	\section{Building and Measurement Models of Interest}\label{Section2}
	The global response of building structures to seismic ground motions can be accurately described for engineering purposes by
	\begin{equation}
	\begin{aligned}
	\mathbf{M}\ddot{q}(t)+\mathbf{C}_{\xi}\dot{q}(t)+f_R(q(t),\dot{q}&(t), z(t))=\\&-\mathbf{M}\mathbf{b}_1\ddot{u}_g(t)+\mathbf{b}_2w(t)
	\label{system}
	\end{aligned}
	\end{equation}
	where the vector $q(t)\in \mathbb{R}^{n}$ contains the relative displacement (with respect to the ground) of all stories. $z(t)$ is a vector of auxiliary variables dealing with material nonlinearity and damage behavior. $n$ denotes the number of geometric DoF, $\mathbf{M}=\mathbf{M}^T \in  \mathbb{R}^{n \times n}$ is the mass matrix, $\mathbf{C}_{\xi}=\mathbf{C}_{\xi}^T \in \mathbb{R}^{n\times n}$ is the damping matrix, $f_R(\cdot) $ is the resultant global restoring force vector. The matrix $\mathbf{b}_1 \in  \mathbb{R}^{n \times r}$ is the influence matrix of the $r$ ground acceleration time histories defined by the vector $\ddot{u}_g(t) \in  \mathbb{R}^{r}$. The matrix $\mathbf{b}_2 \in  \mathbb{R}^{n \times p}$ defines the spatial distribution the vector $w(t) \in \mathbb{R}^p$, which in the context of this paper represents the process noise generated by unmeasured excitations and (or) modeling errors. 
	
	This study relies only on building vibrations measured horizontally in three independent and non-intersecting directions {\color{black} and assumes the vector of acceleration measurements, $\ddot{y}(t) \in \mathbb{R}^{m}$, is given by
		\begin{equation}
		\begin{multlined}[0.5\linewidth]
		\ddot{y}(t) = -\mathbf{c}_2\mathbf{M}^{-1}\biggl[\mathbf{C}_{\xi}\dot{q}(t)+f_R(q(t),\dot{q}(t), z(t))\\\qquad\qquad-\mathbf{b_2}w(t)\biggr]+\nu(t)
		\label{ACC}
		\end{multlined}
		\end{equation}
		where 
		$\mathbf{c}_2 \in  \mathbb{R}^{m \times n}$ is a Boolean matrix that maps the DoFs to the measurements, and $\nu(t) \in \mathbb{R}^{m \times 1}$ is the measurement noise. 
		
		\section{Performance-Based Earthquake Engineering}\label{Section3}
		The aftermath of the 1994 Northridge and 1995 Kobe earthquakes revealed significant vulnerability in the way buildings and other structures were designed to resist earthquakes. Following these, researchers and engineers realized the need to develop seismic design and assessment methods, which can improve the seismic vulnerability of structures and control earthquake losses. These efforts resulted in the development of an important engineering concept known as \textit{performance-based earthquake engineering} (PBEE). PBEE includes concepts and techniques related to the design, construction, and maintenance of structures aimed to ensure (as much as possible) predictable performance objectives are met under earthquake demands. \cite{structural1995performance} In the first generation of the PBEE documents in the United States (also called as PBEE-1), the report of SEAOC Vision 2000 \cite{structural1995performance} made an important step toward the realization of the PBD and PBA of buildings. This report classified the system performance levels as fully operational, operational, life safety, and near collapse, and also, classified hazard levels as frequent, occasional, rare, and very rare events. The stakeholders can determine the desired performance objective of the system based on the system performance levels corresponding to different hazard levels. Subsequent documents of PBEE-1 such as ATC-40 \cite{applied1996seismic}, FEMA-273 \cite{council1997nehrp}, FEMA-356 \cite{fema2000commentary} and ASCE/SEI 41-13 \cite{asce2013seismic} used a similar framework and slightly modified the descriptions for system performance and hazard levels. These documents established approximate relationships between seismic response parameters (inter-story drifts, inelastic element deformations, and element forces) and qualitative performance measures of Immediate Occupancy (IO), Life Safety (LS), Collapse Prevention (CP), and Collapse (C). They also proposed component level acceptance criteria for structural and non-structural elements for various static/dynamic linear/nonlinear analysis. 
		
		The Pacific Earthquake Engineering Research Center (PEER) considered the shortcomings of the PBEE-1 and developed a second-generation of the PBEE framework known as PEER PBEE  (also called as PBEE-2). Figure \ref{fig:peer} presents a summary of the PEER PBEE framework \cite{porter2003overview}; this framework provides a more robust and probabilistic methodology based on four logical steps including 1) hazard analysis, 2) structural analysis, 3) damage analysis and 4) loss analysis.  The outcome of every step is characterized by one of four generalized variables: Intensity Measure (IM), Engineering Demand Parameter (EDP), Damage Measure (DM), and Decision Variable (DV). These variables are defined as follows, {IM} is a parametric representation of ground motion intensity, such as peak ground acceleration, {EDP} is a parametric representation of structural response to ground motion, such as displacements, velocities, accelerations at all degrees of freedom, {DM} is a parametric representation of a damage state such as cracks, failure in connections or structural collapse, and {DV} is a parametric expression of the decision varibale, such as loss expressed in terms of repair costs, casualties or lost occupancy time. Using the Total Probability Theorem, the PEER PBEE framework equation can be expressed
		\begin{equation}
		\begin{aligned}
		p[\text{{DV}}]=\iiint &p[\text{{DV}}|\text{{DM}}]\,p[\text{{DM}}|\text{{EDP}}]\,p[\text{{EDP}}|\text{{IM}}]\text{...}\\&\qquad\qquad\,p[\text{{IM}}|\text{{D}}]
		\,d\text{{IM}}\,.\,d\text{{EDP}}\,.\,d\text{{DM}}
		\end{aligned}
		\end{equation}
		where, the expression $p[\text{X}|\text{Y}]$ refers to the probability density of X conditioned on knowledge of Y; D denotes facility location, structural, non-structural, and other features; $p[\text{IM}|\text{D}]$ is the probability of experiencing a given level of intensity; $p[\text{{EDP}}|\text{{IM}}]$ is the conditional probability of experiencing a level of response, given a level of ground motion intensity; $p[\text{{DM}}|\text{{EDP}}]$ is the conditional probability of experiencing the damage state, given a level of structural response; $p[\text{{DV}}|\text{{DM}}]$ is the conditional probability of experiencing a loss of certain size, given a level of damage. The expected loss or value of the decision variable $p[\text{DV}]$ is calculated as the sum of these quantities over all levels of intensity, response, damage, and loss. Figure \ref{fig:peer} presents a summary of the PEER PBEE framework. The following section aims to present a framework for decision-making regarding the post-earthquake assessment of instrumented buildings in a manner consistent with criteria from performance-based design.
		
		\begin{figure*}
			\centering
			\includegraphics[width=0.9\linewidth]{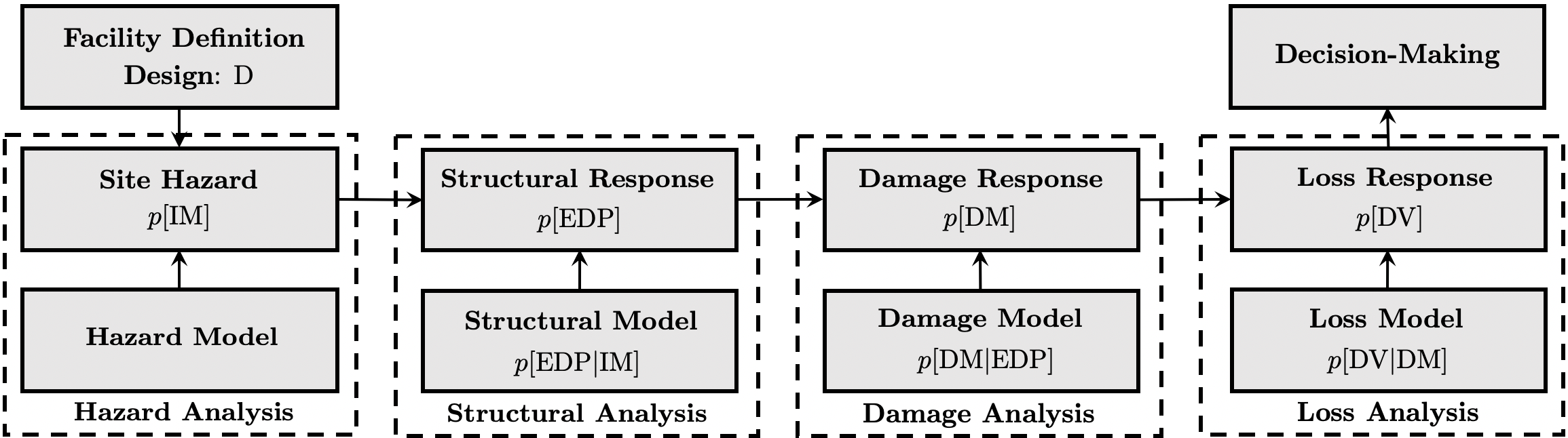}
			\caption{Summary of the PEER PBEE framework}
			\label{fig:peer}
		\end{figure*}

		\section{Proposed Framework for Performance-based Post-earthquake Decision-making}\label{Section4}
		
		The proposed decision-making framework stems from the performance-based design framework and consists of the following four steps: 
		1) optimal sensor placement, 2) response reconstruction, 3) damage estimation, and 4) performance-based assessment and decision-making. Figure \ref{fig:pbm} presents a summary of the proposed framework, where the outcome of every step of the proposed framework is characterized by one of four generalized variables,  Response Measurement (M), Engineering Demand Parameter (EDP), Damage Measure (DM), and Decision Variable (DV); where M is parametric representation of measurement intensity. Using the Total Probability Theorem, the proposed framework equation is expressed by
		\begin{equation}
		\begin{aligned}
		p[\text{{DV}}]=\iiint &p[\text{{DV}}|\text{{DM}}]\,p[\text{{DM}}|\text{{EDP}}]\,p[\text{{EDP}}|\text{{M}}]\text{...}\\&\,\,\quad\qquad\qquad\,p[\text{{M}}]\,d\text{{M}}\, . \,d\text{{EDP}}\, . \,d\text{{DM}}
		\label{PBMeq}
		\end{aligned}
		\end{equation}
		where $p[\text{M}]$ is the probability density of measurement set, and  $p[\text{{EDP}}|\text{{M}}]$ is the conditional probability of experiencing a level of response given measurement set $\text{M}$. Except a few special cases, solving the multidimensional integrals in Equation \ref{PBMeq} is very complex and challenging task as it requires the complete probability distribution of each three generalized variables (DM, EDP and DM) to be estimated. For instance, to estimate $p[\text{{EDP}}|\text{{M}}]$ in the special case of linear structural systems (which can be described by linear models), the densities $p[\text{{EDP}}|\text{{M}}]$ are Gaussian. This means they can be characterized by mean vectors and covariance matrices; thus, the mathematical solution becomes trackable. This is important because in real world application there are many cases that can be addressed using this special case. However, in the case of more complicated systems, where there is a need to solve the nonlinear filtering problem, there does not exist a finite set of parameters that can characterize the densities $p[\text{{EDP}}|\text{{M}}]$. Instead, we seek algorithms that can provide estimates based on approximations of the probability density functions using the first two statistical moments \cite{roohi2019performance}. In the following, each step of soliving the Equation \ref{PBMeq} is discussed in more details to obtain approximate solution of the $p[\text{DV}]$ and use the outcome for post-earthquake decision-making.
		\begin{figure*}
			\centering
			\includegraphics[width=0.9\linewidth]{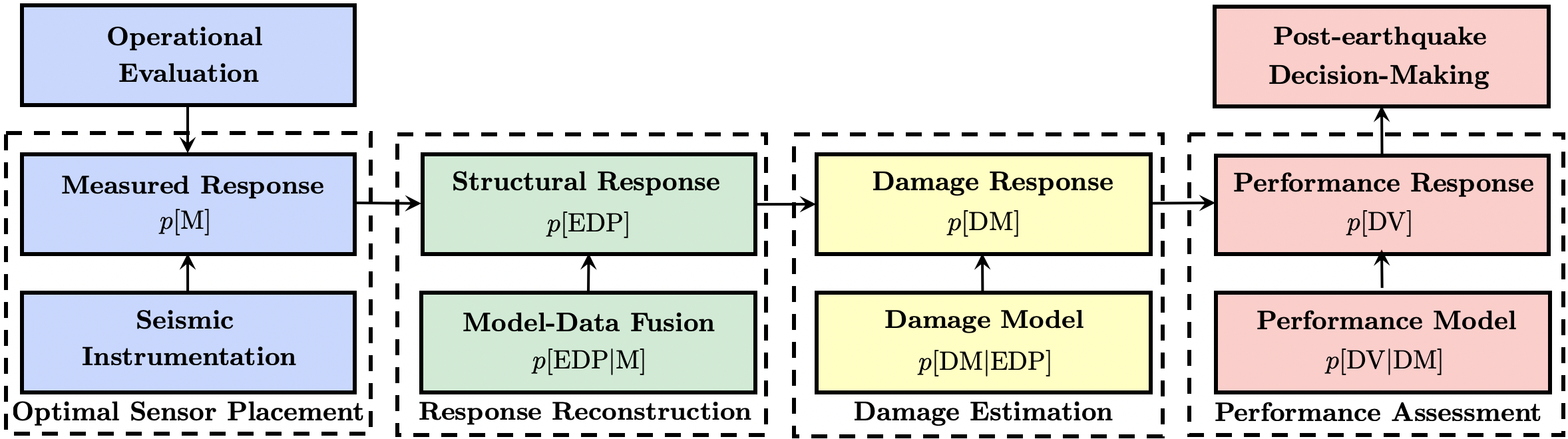}
			\caption{Summary of the proposed performance-based post-earthquake decision-making framework}
			\label{fig:pbm}
		\end{figure*}
		%
		%
		%
		
		\subsection{Optimal Sensor Placement}
		The first step of the proposed framework is to determine the type, number, and location of the sensors. In practice, this process begins with sensor type selection considering technical, logistical, and economic constraints. This paper is restricted to accelerometers due to their popularity in typical seismic instrumentation. Thus, this step requires only to determine the number and locations of the sensors, which is typically known as \textit{\say{optimal sensor placement}} problem. Here, the meaning of the term \textit{\say{optimal}} depends on the objective of sensor placement, which can be identification, damage detection, or response reconstruction. The aim is to place accelerometers in locations that contain maximum information for response reconstruction, i.e., select the number and locations of sensors in a way that minimizes the uncertainty of response reconstruction. This minimization can be achieved by selecting an optimality criterion based on the variance of a user-defined objective function related to the state of the system, such as displacement, internal forces, and stresses. The proposed framework selects the objective function to be the sum of displacement estimation error variances or the sum of diagonal elements of displacement estimation error covariance matrix. Therefore, an optimal sensor placement can be achieved by solving an optimization problem to select the optimal $\mathbf{c}_2$ (in Equation \ref{ACC}) subject to 
		maximum inter-story drift (ISD) estimation variance being bounded by a maximum allowable variance of $\sigma^2_{\mathrm{max}}$, which can be specified based on the expected accuracy to determine performance-based post-earthquake re-occupation category of the building of interest. This optimization problem can be formulated as follows
		\begin{equation}\label{c2}
		\begin{aligned}
		(\mathbf{c}_2)_{opt}     =&  \argminA_{\mathbf{c}_2}  \, tr(\mathbf{P}) \\
		\text{s.t.}\quad&     \mathrm{max}\bigl[\sigma^2_{\text{ISD}_{(k,k)}}\bigr]_{k=1:n}<\sigma^2_{\mathrm{max}}
		\end{aligned}
		\end{equation}
		where $k$ is story number, $n$ is total number of stories, $tr$ is trace of a square matrix (defined as the sum of elements on the  diagonal), $\mathbf{P}$ is displacement estimation error covariance matrix given by
		\begin{equation}
		\mathbf{P}=\mathbb{E}\left[[q(t)-\hat{q}(t)][q(t)-\hat{q}(t)]^T\right]
		\end{equation}
		$\hat{q}(t)$ is displacement estimate, and $\sigma^2_{\text{ISD}_{(k,k)}}$ is the $k$th diagonal element of inter-story estimation error covariance matrix, $\mathbf{{P}}_{\text{ISD}_{(k,k)}}$, given by
		\begin{equation}
		\begin{aligned}
		\lefteqn{\mathbf{{P}}_{\text{ISD}_{(k,k)}} =}\\
		& \qquad\left\{\begin{array}{ll}
		\mathbf{P}_{(1,1)} &\text{for }k=1\\
		\mathbf{P}_{(k,k)}+\mathbf{P}_{(k-1,k-1)}-2\mathbf{P}_{(k,k-1)} &\text{for }k\neq1
		\end{array}\right.
		\end{aligned}
		\end{equation}
		Section \ref{nmbo}, proposes an expression to determine $\mathbf{P}$ (see Equation \ref{Pequation}).  
		
		\subsection{Response Reconstruction}
		Once data becomes available from a seismic event, given by $p[\text{M}]$, response reconstruction is the second step of the proposed framework. Response reconstruction refers to the estimation of unmeasured response quantities of interest or engineering demand parameters (EDP) from a limited number of global response measurements, given by $p[\text{EDP}|\text{M}]$. The information needed for reconstructing seismic response are the following: 1) the dynamic response of the building at all DoF and 2) a mapping between the global and local DoF of every element. An accurate response reconstruction in the step is vital to prevent under-estimation or over-estimation of the actual response of the building. Further, the estimated uncertainty bound helps to develop a set of maximum, mean, and minimum seismic demand to consider the best and worst-case scenarios in assessing the performance of the instrumented building.
		
		In the literature, researchers have proposed four categories of state observers to perform response reconstruction based on sub-optimal nonlinear filters including: classical nonlinear Bayesian filters (e.g., \textit{extended Kalman filter} (EKF) \cite{gelb1974applied}), modern nonlinear Bayesian (or statistically linearized) filters (e.g., \textit{unscented Kalman filter} (UKF) \cite{julier2000new}, particle-based nonlinear Bayesian filters (e.g. the \textit{particle filter} \cite{doucet2000sequential}), and nonlinear model-data fusion using state observers (e.g., \textit{nonlinear model-based observer} (NMBO) \cite{Roohi2019nonlinear}). From these response reconstruction approaches, the proposed framework uses the NMBO for response reconstruction in instrumented buildings. This is mainly because the author expects that the use of better modeling capabilities will significantly improve the accuracy of response reconstruction. The estimated response parameters with their associated uncertainties can form a demand set to perform damage estimation. 
		
		\subsubsection{Nonlinear model-based observer}\label{nmbo}
		The NMBO estimate of the displacement response, ${\hat{q}}(t)$, is given by the solution of the following set of ordinary differential equations
		\begin{equation}\label{NMBO}
		\begin{multlined}[0.5\linewidth]
		\mathbf{M}\ddot{\hat{q}}(t)+(\mathbf{C}_{\xi}+\mathbf{c}_{2}^{T}\mathbf{E}\mathbf{c}_{2})\dot{\hat{q}}(t)+f_R(\hat{q}(t),\dot{\hat{q}}(t),z(t))\\=\mathbf{c}_{2}^{T}\mathbf{E}\dot{y}(t)
		\end{multlined}
		\end{equation}
		where $\dot{y}(t)$ is the measured velocity and $\mathbf{E}\in  \mathbb{R}^{m \times m}$ is the feedback gain. It can be seen that Equation \ref{NMBO} is of the same form of the original nonlinear model of interest in Equation \ref{system}. A physical interpretation of the NMBO can be obtained by viewing the right-hand side of Equation \ref{NMBO} as a set of corrective forces applied to a modified version of the original nonlinear model of interest in the left-hand side. The modification consists in adding the damping term $c_2^T\mathbf{E}c_2$, where the matrix $\mathbf{E}$ is free to be selected. The diagonal terms of $\mathbf{E}$ are equivalent to grounded dampers in the measurement locations, and the off-diagonal terms (typically set to zero) are equivalent to dampers connecting the respective DoF of the measurement locations. To retain a physical interpretation, the constraints on $\mathbf{E}$ are symmetry and positive definiteness. Also, the corrective forces $\mathbf{c}_{2}^{T}\mathbf{E}\dot{y}(t)$ are proportional to the velocity measurements and added grounded dampers. The velocity measurements $\dot{y}(t)$ can be obtained by integration of acceleration measurements $\ddot{y}(t)$ in Equation \ref{ACC}. The integration might add long period drifts in velocity measurements, and high-pass filtering can be performed to remove these baseline shifts. To determine $\textbf{E}$, the objective function to be minimized is the trace of the estimation error covariance matrix. Since for a general nonlinear multi-variable case, a closed-form solution for the optimal matrix $\mathbf{E}$ has not been found, a numerical optimization algorithm is used. To derive the optimization objective function, Equation \ref{NMBO} is linearized as follows
		\begin{equation}\label{LNMBO}
		\mathbf{M}\ddot{\hat{q}}(t)+(\mathbf{C}_{\xi}+\mathbf{c}_{2}^{T}\mathbf{E}\mathbf{c}_{2})\dot{\hat{q}}(t)+\mathbf{K}_0{\hat{q}}(t)=\mathbf{c}_{2}^{T}\mathbf{E}\dot{y}(t)
		\end{equation}where  $\mathbf{K}_0$ is the initial stiffness matrix. By defining the state error as $e=q-\hat{q}$, {\color{black}it was shown in \cite{hernandez2011natural} that} the PSD of estimation error, $\pmb{\Phi}_{ee}$, is given by
		\begin{equation}\label{Phie}
		\begin{aligned}
		\pmb{\Phi}_{ee}(\omega)=&\, \mathbf{H}_o\, \mathbf{b}_2 \, \pmb{\Phi}_{ww}(\omega)\, \mathbf{b}_2^{T}\, \mathbf{H}_o^*+ \text{...}\\
		&\quad\qquad\qquad\qquad\mathbf{H}_o\, \mathbf{c}_2^T\, \mathbf{E} \, \pmb{\Phi}_{vv}(\omega)\, \mathbf{E}^T\, \mathbf{c}_2\, \mathbf{H}_o^*
		\end{aligned}
		\end{equation}
		with $\mathbf{H}_o$ defined as
		\begin{eqnarray}
		\mathbf{H}_o=\left(-\mathbf{M}\omega^2+\left(\mathbf{C}_{\xi}+\mathbf{c}_2^T\mathbf{Ec}_2\right)i\omega+\mathbf{K}_0\right)^{-1}
		\end{eqnarray}
		where the matrices $\pmb{\Phi}_{ww}(\omega)$ and $\pmb{\Phi}_{vv}(\omega)$ are the PSDs of the uncertain excitation on the system and measurement noise, respectively. {\color{black} In this paper, the uncertain input corresponds to the ground motion excitation, and the measurement noise corresponds to unmeasured excitations and (or) modeling errors}. To select the optimal value of $\mathbf{E}$ matrix, the following optimization problem must be solved
		\begin{equation}\label{J}
		\begin{aligned}
		\mathbf{E}_{{opt}} = &  \argminA_{\mathbf{E}}  \, tr(\mathbf{P}) \\
		\text{s.t.}\quad&{\mathbf{E} \, \in \, \mathbb{R}^{+}}
		\end{aligned}
		\end{equation}
		where $\mathbf{P}$ is the covariance matrix of displacement estimation error described as
		\begin{equation}
		\begin{aligned}
		\mathbf{P}&=\mathbb{E}\biggl[\bigl[q(t)-\hat{q}(t)][q(t)-\hat{q}(t)\bigr]^T\biggr]
		\\&=\int_{-\infty}^{+\infty}\pmb{\Phi}_{ee}(\omega)d\omega
		\label{Pequation}
		\end{aligned}
		\end{equation}
		One alternative for the optimization problem (in Equation \ref{J}) can be defined if the objective is minimization of the inter-story drifts estimation error covariance matrix by formulating the optimization problem based on $\mathbf{P_\text{ISD}}$ as follows
		\begin{equation}\label{JT}
		\begin{aligned}
		\mathbf{E}_{{opt}} =& \argminA_{\mathbf{E}}  \, tr(\mathbf{P_\text{ISD}}) \\
		\text{s.t.}\quad&{\mathbf{E} \, \in \, \mathbb{R}^{+}}
		\end{aligned}
		\end{equation}
		where 
		\begin{equation}
		\begin{aligned}
		&tr(\mathbf{P_\text{ISD}}) = \sum_{k=1}^n\mathbf{P_\text{ISD}}_{(k,k)} \\
		&=\,\mathbf{P}_{(1,1)}+\sum_{k=2}^n[\mathbf{P}_{(k,k)}+\mathbf{P}_{(k-1,k-1)}-2\mathbf{P}_{(k,k-1)}]
		\end{aligned}
		\end{equation}
		Any optimization algorithm (e.g., Matlab \say{\textit{fminsearch}}) can be used to solve the optimization in Equations \ref{J} and \ref{JT} by varying the values of the diagonal elements of the $\mathbf{E}$ matrix to determine the optimized feedback matrix.  Figure \ref{fig:nmbox} presents a summary of the nonlinear model-data fusion using the NMBO. Also, readers are kindly referred to \cite{hernandez2011natural,hernandez2013optimal,hernandez2018estimation,roohi2019nonlinearStanford}
		for implementation examples.} 
	\begin{figure}
		\centering
		\includegraphics[width=1\linewidth]{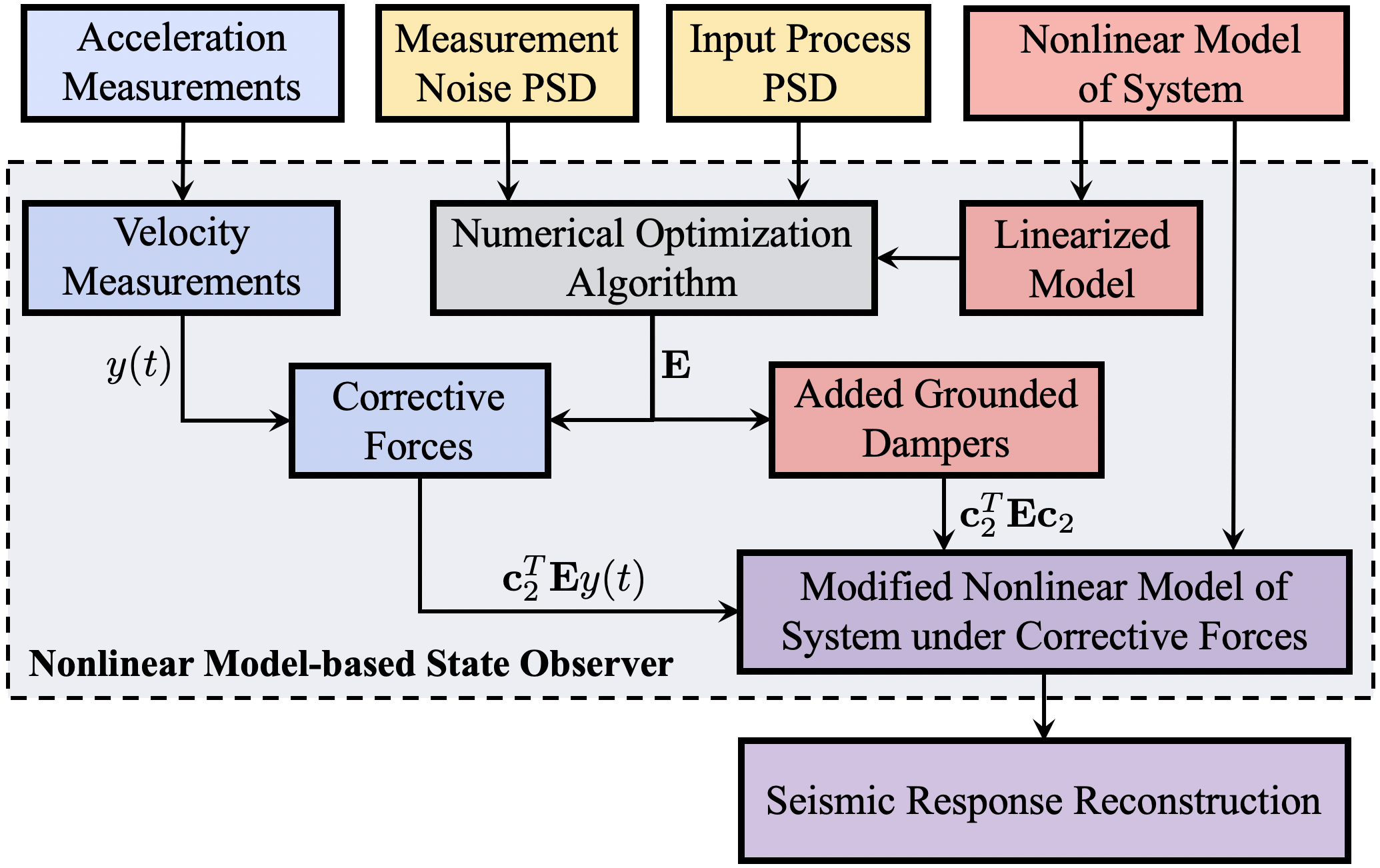}
		\caption{Summary of the nonlinear model-data fusion using the nonlinear model-based observer (NMBO)}
		\label{fig:nmbox}
	\end{figure}
	
	\subsection{Damage Estimation}   \index{Performance-Based Seismic Monitoring!Damage Estimation}  
	The third step of the proposed framework is to estimate damage measure (DM) from the estimated response and compare the DMs with performance-based acceptance criteria. The outcome of this step is given by $p[\text{DM}|\text{EDP}]$, which is the probability of DM given EDP. Based on the selected damage measure, the $p[\text{DM}|\text{EDP}]$ is calculated at the element or system level. Then, the outcome is evaluated using the acceptance criteria to determine the post-earthquake re-occupancy category of the instrumented building and also to detect and localize element-level structural damage. Based on the observations from past earthquakes that the main portion of the seismic damage and loss to the structural and non-structural elements are associated with excessive geometric deformations such as inter-story drifts. Therefore, the maximum inter-story drift is considered as damage measure, and the $p[\text{DM}]$ ($=p[\text{ISD}]$) is reconstructed from the estimated EDPs as discussed in the following.
	
	\subsubsection{Inte-story Drift Estimation}
	The expected value of {maximum} ISD estimate at each story can be calculated using the NMBO displacement estimates ($\hat{q}(t)$) as follows      
	\DeclarePairedDelimiter\abs{\lvert}{\rvert}
	\newcommand{\Mypm}{\mathbin{\tikz [x=1.4ex,y=1.4ex,line width=.1ex] \draw (0.0,0) -- (1.0,0) (0.5,0.08) -- (0.5,0.92) (0.0,0.5) -- (1.0,0.5);}}%
	\begin{eqnarray}\label{Drift}
	\mathbb{E}\bigl[\text{ISD}_{{k}}\bigr]=\frac{\mathrm{max}\abs[\Big]{\hat{q}_k(t) - \hat{q}_{k-1}(t)}}{h_k}
	\end{eqnarray}
	where $h_k$ is height of $k$-th story, and the uncertainty in ISD estimation can be calculated as follows 
	\begin{eqnarray}\label{Driftsigma}
	\sigma^2_{{\text{ISD}_k}}={\mathbf{{P}}_{\text{ISD}_{(k,k)}}}
	\end{eqnarray}
	where $\sigma_{{\text{ISD}_k}}$ is the uncertainty standard deviation of ISD estimation for $k$-th story. The estimated ISDs and their uncertainties are subsequently used to reconstruct probability density function of ISD for each story, $p(\text{ISD}_k)$, assuming a Gaussian (normal) distribution as follows 
	\begin{equation}
	p(\text{ISD}_k)\sim \mathcal{N}(\mathbb{E}\bigl[\text{ISD}_{{k}}\bigr],\,\sigma_{{\text{ISD}_k}}^{2})\,
	\end{equation}

	\subsection{Post-earthquake Reoccupancy Classification and Decision-Making}
	The estimated $p[\text{DM}]$ is subsequently used as input to performance model, $p[{\text{DV}}|{\text{DM}}]$ to estimate $p[\text{DV}]$ or $p[\text{DV}\geq\text{PL}]$, definded as probability of DV exceeding specific performance level (PL) based on performance-based acceptance criteria. The acceptance criteria relate engineering demand parameters (such as inter-story drifts, inelastic element deformations, and element forces) to qualitative performance levels of Immediate Occupancy (IO), Life Safety (LS), Collapse Prevention (CP), and Collapse (C) \cite{FEMA-356}. Therefore, the probability of exceeding specific perfomance level (PL), $p[\text{ISD}_k\geq\text{PL}] $, can be calculated for each story as follows 
	\begin{equation}
	\begin{aligned}
	p[\text{DV}_k\geq\text{PL}] &= \int_\text{PL} p[{\text{DV}_k}|{\text{DM}_k}] \,d\text{DM}_k\\
	&= \int_{\text{PL}}^{} p(\text{ISD}_k) \,d\text{ISD}_k\\&= F_{\text{ISD}}(\text{ISD}_k\geq{\text{PL}})
	\end{aligned}
	\end{equation}
	where $p(\text{ISD}_k\geq\text{PL})$ is the probability of $\text{ISD}_k$ exceeding specific performance levels (PL) at story $k$, and $F_{\text{ISD}}$ is the cumulative probability density (CDF) of estimated ISD at story $k$. Here, performance levels include IO, LS, CP and C. Additionally, the probability of specific performance level, $p[\text{ISD}_k=\text{PL}]$, can be obtained for four classes of performance levels including $\text{ISD}_k<\text{IO}$, $\text{ISD}_k\geq\text{IO}$, $\text{ISD}_k\geq\text{LS}$, and $\text{ISD}_k\geq\text{CP}$ as follows
	\begin{equation}
	\begin{aligned}
	\lefteqn{p[\text{DV}_k=\text{PL}] =}\\
	&\left\{\begin{array}{ll}
	1-F_{\text{ISD}}(\text{ISD}_k\geq\text{IO}) & \,\,\,\,\text{for}\,\, \text{PL} = \text{IO}\\
	F_{\text{ISD}}(\text{ISD}_k\geq\text{LS})-F_{\text{ISD}}(\text{ISD}_k\geq\text{IO})& \,\,\,\,\text{for}\,\, \text{PL} = \text{LS}\\
	F_{\text{ISD}}(\text{ISD}_k\geq\text{CP})-F_{\text{ISD}}(\text{ISD}_k\geq\text{LS})&\,\,\,\,\text{for}\,\, \text{PL} = \text{CP}\\
	F_{\text{ISD}}(\text{ISD}_k\geq\text{CP})&\,\,\,\,\text{for}\,\, \text{PL} = \text{C}
	\end{array}\right.
	\end{aligned}
	\end{equation}
	Then, the post-earthquake building classification can be determined based on the probabilities obtained from inter-story performance asssessmenet. Assuming that the $\text{ISD}$s are independent, the probability of exceeding specific performance level, $p[\text{ISD}\geq\text{PL}]$, for the building can be calculated as follows
	\begin{equation}
	\begin{aligned}
	p[\text{DV}\geq\text{PL}]&=p[\text{ISD}\geq\text{PL}]\\
	&=1-\prod_{k=1}^n (1-p[\text{ISD}_k\geq\text{PL}])\\
	&=1-\prod_{k=1}^n (1-F_{\text{ISD}}[\text{ISD}_k\geq\text{PL}])
	\end{aligned}
	\end{equation}
	and similarly, building-level probability of specific performance level, $p[\text{ISD}=\text{PL}]$, can be obtained for four classes of performance levels as follws
	\begin{equation}
	\begin{aligned}
	\lefteqn{p[\text{DV}=\text{PL}] =}\\
	&\left\{\begin{array}{ll}
	1-p(\text{ISD}\geq\text{IO}) & \,\,\,\,\text{for}\,\, \text{PL} = \text{IO}\\
	p(\text{ISD}\geq\text{LS})-p(\text{ISD}\geq\text{IO})& \,\,\,\,\text{for}\,\, \text{PL} = \text{LS}\\
	p(\text{ISD}\geq\text{CP})-p(\text{ISD}\geq\text{LS})&\,\,\,\,\text{for}\,\, \text{PL} = \text{CP}\\
	p(\text{ISD}\geq\text{CP})&\,\,\,\,\text{for}\,\, \text{PL} = \text{C}
	\end{array}\right.
	\end{aligned}
	\end{equation}

	\section{Case-study: Van Nuys Hotel Testbed}
	This section illustrates the proposed framework using seismic response measurements from Van Nuys hotel. The CSMIP instrumented this building as Station 24386, and the recorded data of this building are available from several earthquakes, including 1971 San Fernando, 1987 Whittier Narrows, 1992 Big Bear, and 1994 Northridge earthquakes. From these data, measurements during 1992 Big Bear and 1994 Northridge earthquakes are used in this study to demonstrate the proposed framework. Researchers have widely studied the Van Nuys building, and the building was selected as a testbed for research studies by researchers in PEER \cite{Krawinkler2005}. 
	
	\subsection{Description of the Van Nuys building, Building instrumentation, and Earthquake damage}
	The case-study building is a 7-story RC building located in San Fernando Valley in California. The building plan is 18.9 m $\times$ 45.7 m in the North-South and East-West directions, respectively. The total height of the building is 19.88 m, with the first story of 4.11 m tall, while the rest are 2.64 m approximately. The structure was designed in 1965 and constructed in 1966. Its vertical load transfer system consists of RC slabs supported by concrete columns and spandrel beams at the perimeter. The lateral resisting systems are made up of interior concrete column-slab frames and exterior concrete column-spandrel beam frames. The foundation consists of friction piles, and the local soil conditions are classified as alluvium. The testbed building is described in more detail in \cite{Trifunac1999,Krawinkler2005}.
	
\begin{figure}
		\centering
		\subfigure[Van Nuys hotel testbed]{\includegraphics[width=0.7\linewidth]{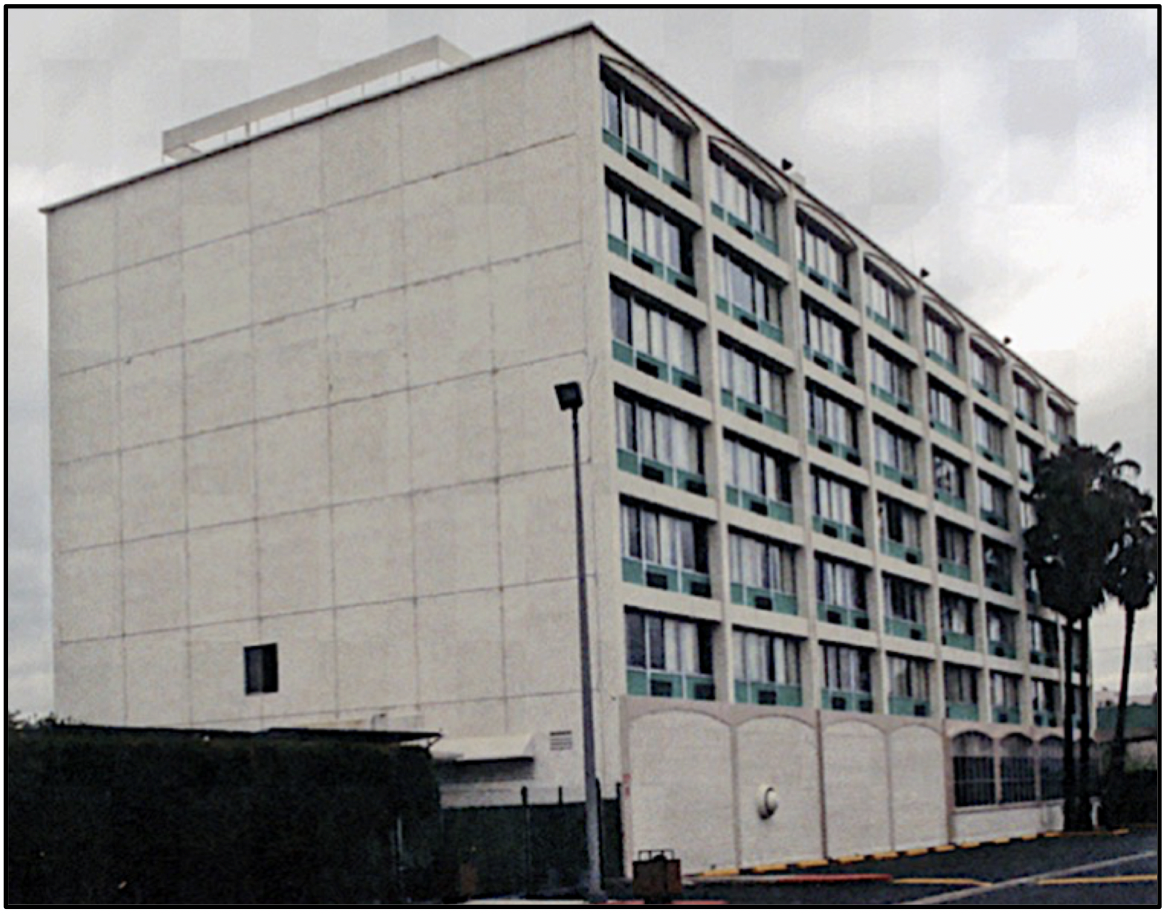}}\\
		\subfigure[Building instrumentation]{\includegraphics[width=1\linewidth]{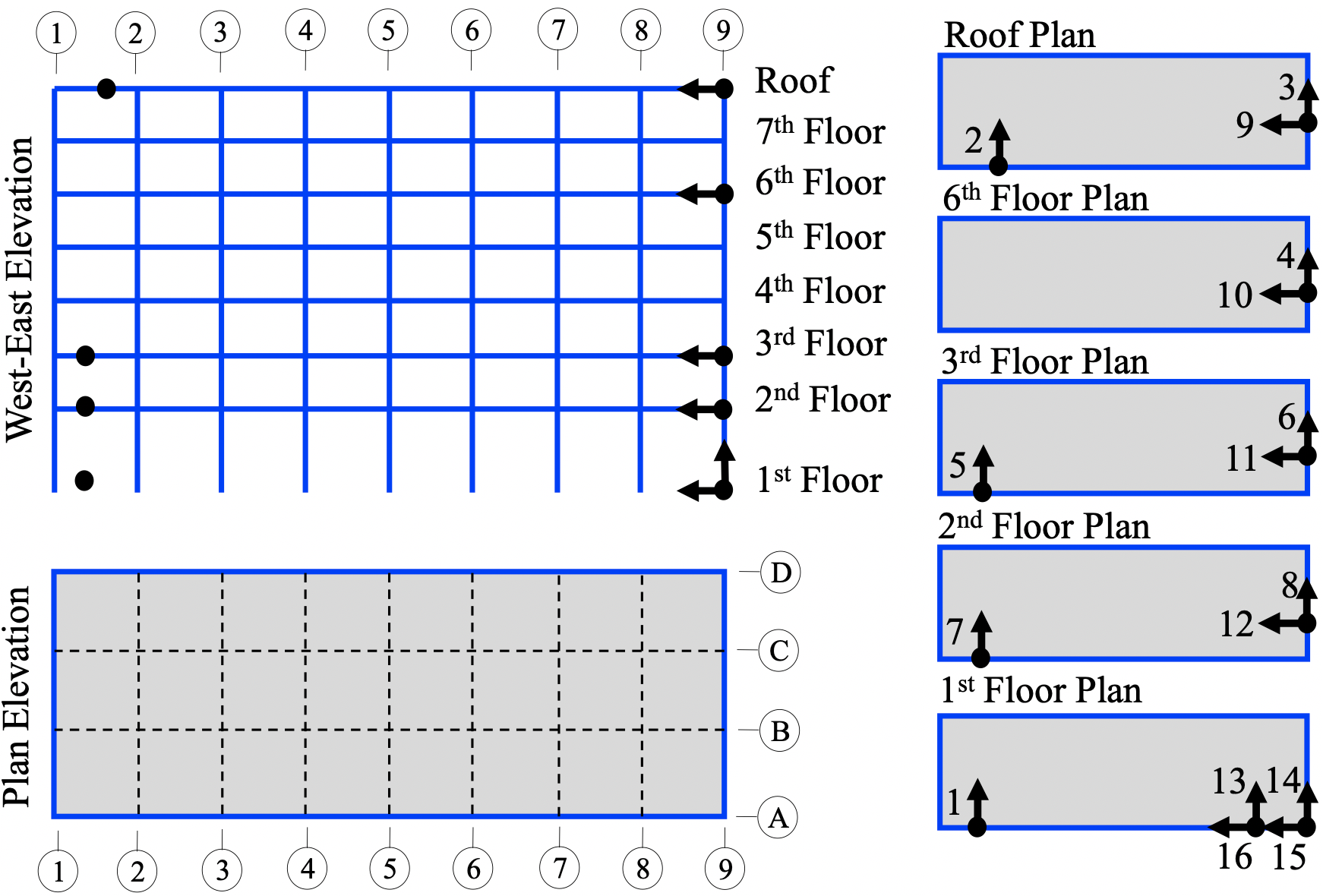}}
		\vspace{-5pt}
		\caption{(top) Van Nuys hotel testbed (CSMIP Station 24386) and (bottom) Location of building accelerometers on the West-East elevation and floor plans}
		\label{fig:Sensors}
	\end{figure}

	The CSMIP initially instrumented the building with nine accelerometers at the 1st, 4th, and roof floors. Following the San Fernando earthquake, CSMIP replaced the recording layout by 16 remote accelerometer channels connected to a central recording system. These channels are located at 1st, 2nd, 3rd, 6th, and roof floors. Five of these sensors measure longitudinal accelerations, ten of them measure transverse accelerations, and one of them measures the vertical acceleration. Figure \ref{fig:Sensors} shows the location of accelerometers.
	
	Since the Van Nuys building was instrumented and inspected following earthquakes that affected the structure, the history of damage suffered by this building is well-documented. These documents show that the building has experienced insignificant structural and mostly nonstructural damage before the Northridge earthquake in 1994. However, the Northridge earthquake extensively damaged the building. Post-earthquake inspection red-tagged the building and revealed that the damage was severe in the south longitudinal frame (Frame A). In Frame A, five of the nine columns in the 4th story (between floors 4 and 5) were heavily damaged due to inadequate transverse reinforcement, and shear cracks ($\geq 5cm$) and bending of longitudinal reinforcement were easily visible \cite{Trifunac2003}.

	%
	
	
	\subsection{Response Reconstruction}
	A nonlinear finite element (FE) model and response measurements of the Van Nuys building was employed to implement the NMBO in OpenSEES and perform response reconstruction. The following subsections present the step-by-step formulation of the OpenSEES-NMBO.
	
	\subsubsection{Nonlinear modeling of the Van Nuys hotel testbed in OpenSEES}
	The nonlinear FE model of the building was implemented using a two-dimensional fixed-base model within the environment of OpenSEES \cite{opensees}. This model corresponds to one of the longitudinal frames of the building (Frame A in Figure \ref{fig:Sensors}). In the FE model, beams and columns were modeled based on distributed plasticity modeling approach, and the \textit{force-based beam-column} elements were used to accurately determine yielding and plastic deformations at the integration points along the element. Gauss-Lobatto integration approach was employed to evaluate the nonlinear response of force-based elements. Each beam and column element was discretized with four integration points, and the cross-section of each element was subdivided into fibers. The uniaxial \textit{Concrete01} material was selected to construct a \textit{Kent-Scott-Park} object with a degraded linear unloading and reloading stiffness and zero tensile strength. The uniaxial \textit{Steel01} material was used to model longitudinal reinforcing steel as a bilinear model with kinematic hardening. The elasticity modulus and strain hardening parameters were assumed to be 200 GPa and 0.01, respectively. Due to insufficient transverse reinforcement in beams and columns \cite{Jalayer2017}, an unconfined concrete model was defined to model concrete. The peak and post-peak strengths were defined at a strain of 0.002 and a compressive strain of 0.006, respectively. The corresponding strength at ultimate strain was defined as $0.05f'_c$ for $f'_c= 34.5$ MPa and $f'_c= 27.6$ MPa and $0.2f'_c$ for $f'_c= 20.7$ MPa. Based on the recommendation of \cite{Islam1996}, the expected yield strength of Grade 40 and Grade 60 steel were defined as 345 MPa (50 ksi) and 496 MPa (72 ksi), respectively, to account for inherent overstrength in the original material and strength gained over time.
	
	\subsubsection{PSD selection and numerical optimization}
	The PSD of ground motion, $\pmb{\Phi}_{ww}(\omega)$, was characterized using the Kanai-Tajimi PSD given by
	\begin{equation}\label{KT}
	S(\omega)=G_0\frac{1+4\xi_g^{2}(\frac{\omega}{\omega_g})^{2}}{\left[1-(\frac{\omega}{\omega_g})^2\right]^2+4\xi_g^{2}(\frac{\omega}{\omega_g})^{2}}
	\end{equation} 
	and the amplitude modulating function $I(t)$ was selected as
	\begin{equation}\label{It}
	I(t)=te^{-\alpha{t}}
	\end{equation}
	The parameter were defined as $\xi_g=0.35$ for both earthquakes, $\omega_g=6\pi rad/s$ for Northridge earthquake and $\omega_g=2\pi rad/s$ for Big Bear earthquake. The underlying white noise spectral density $G_0$ for each direction of measured ground motion for each shake table test was found such that about 95\% of the Fourier transform of the measured ground motion lies within two standard deviations of the average from the Fourier transforms of an ensemble of 200 realizations of the Kanai-Tajimi stochastic process. $\alpha$ was selected as 0.12. Details can be found in \cite{Roohi2019nonlinear}. Also, the PSD of measurement noise, $\pmb{\Phi}_{vv}(\omega)$, in each measured channel was taken as zero mean white Gaussian sequences with a noise-to-signal root-mean-square (RMS) ratio of 0.02. Numerical optimization was performed using Equation \ref{JT}. Table \ref{table:assembly} presents the optimized damper values for each seismic event.
	
	\begin{table}
		\caption{Optimized damper values in kN.s/m (kips.s/in) units}
		\label{table:assembly}
		\centering
		\small
		\renewcommand{\arraystretch}{1.25}
		\begin{tabular}{l c l c l c |}
			\hline 
			Story & Big Bear earthquake & Northridge earthquake\\
			\hline
			1 & 7283.11 (41.59) & 5209.72 (29.75)\\
			2 & 9357.25 (53.43) & 6592.45 (37.64)\\
			5 & 19299.40 (110.20) & 16612.79 (94.86)\\
			7 & 34808.04 (198.76) & 47217.69 (269.62)\\
			\hline 
		\end{tabular}
		\normalsize
	\end{table}
	
	\subsubsection{Formulation of the OpenSEES-NMBO}
\begin{figure}
		\centering
		\subfigure[Van Nuys hotel testbed]{\includegraphics[width=1\linewidth]{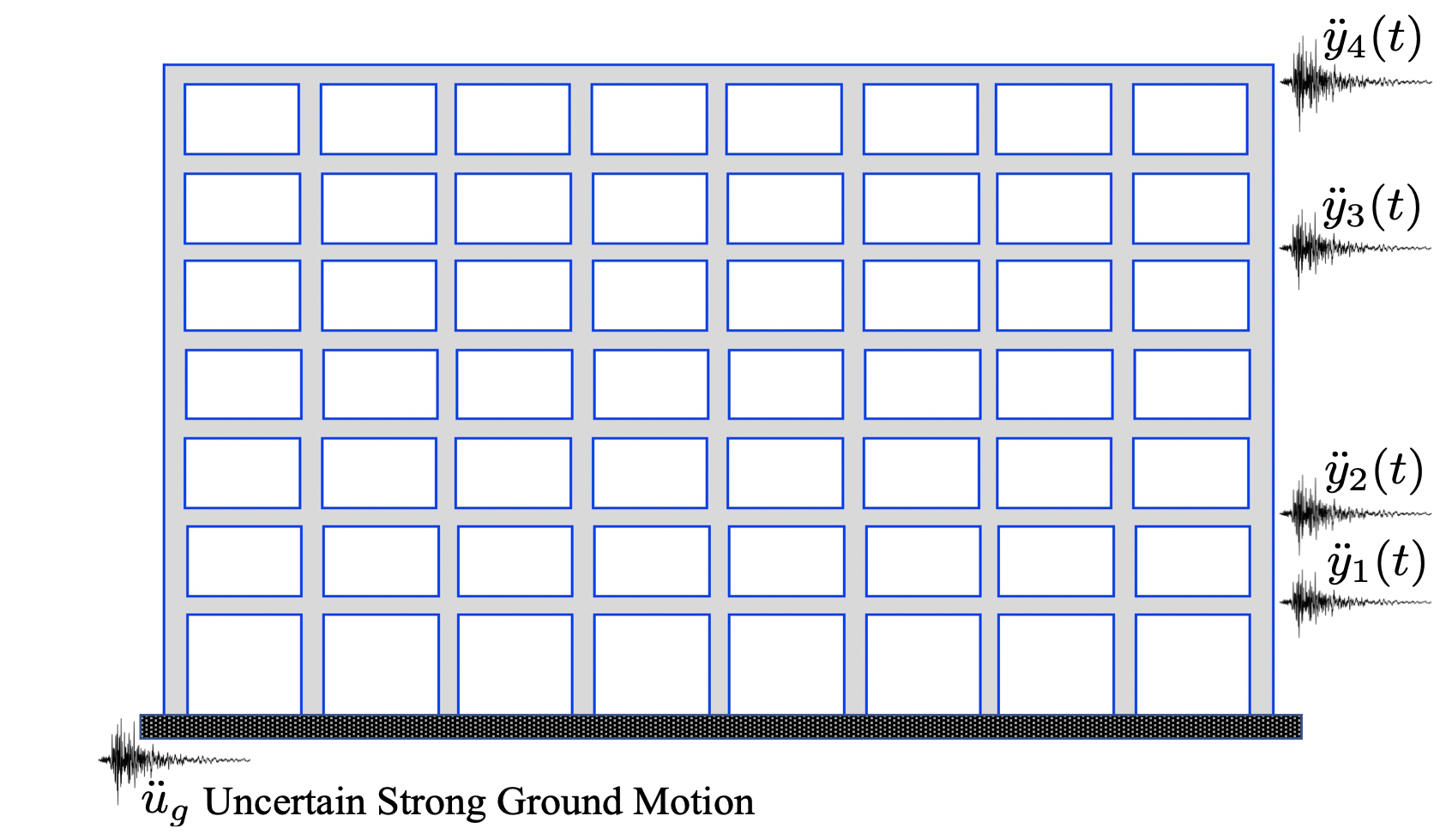}}\\
		\vspace{-10pt}
		\subfigure[OpenSEES-NMBO]{\includegraphics[width=1\linewidth]{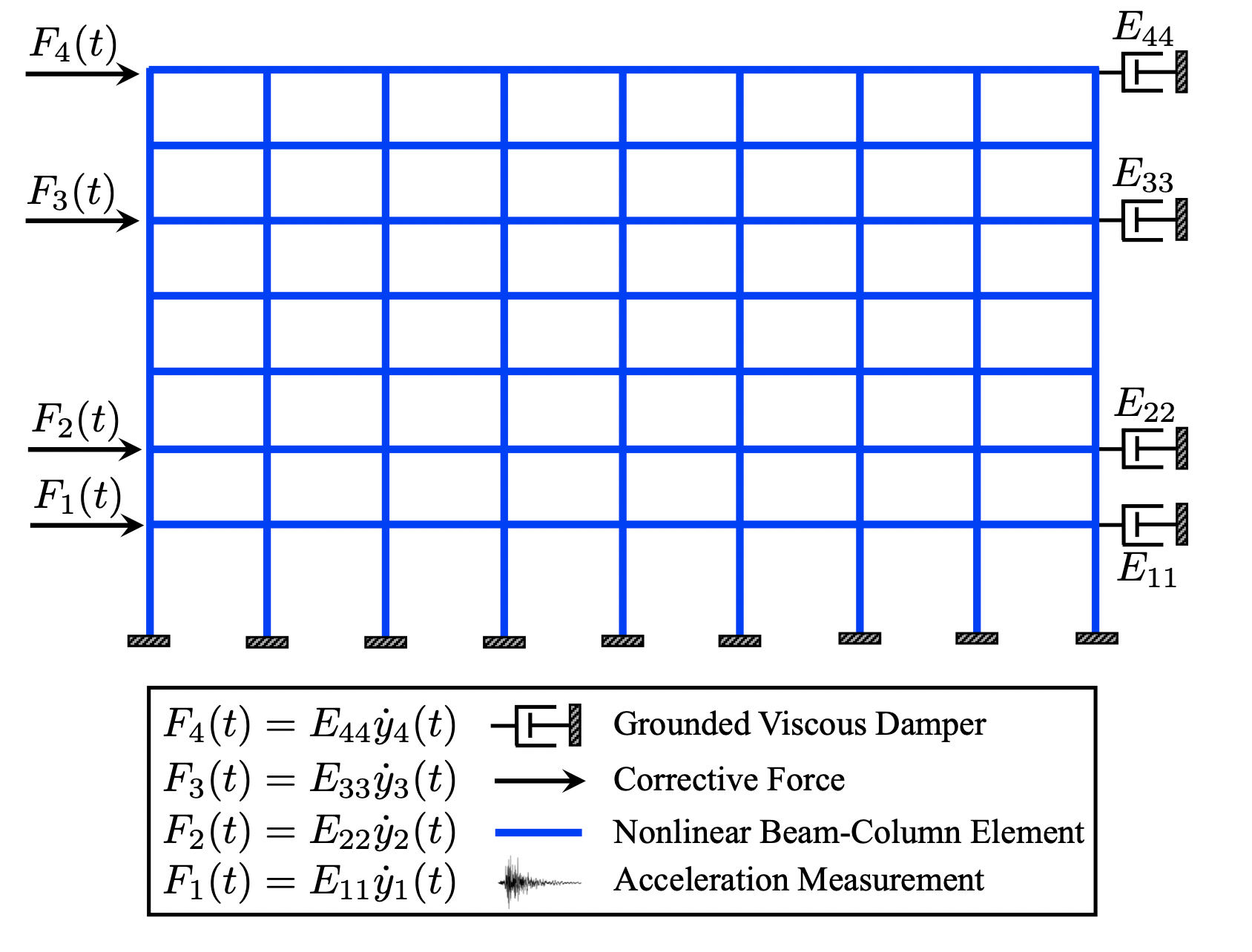}}
		\vspace{-5pt}
		\caption{Schematic of the Van Nuys hotel testbed with location of accelerometers (left) and the OpenSEES-NMBO with corresponding added viscous dampers and corrective forces in measurement locations}
		\label{vannuysnmbo}
	\end{figure}

	The OpenSEES nonlinear FE model was modified by adding grounded dampers in measurement locations and was subjected to corrective forces. Dynamic analysis was performed to estimate the complete seismic response. Figure \ref{vannuysnmbo} presents a schematic of the Van Nuys hotel testbed (with the location of accelerometers) along with the OpenSEES-NMBO (with corresponding added viscous dampers and corrective forces in measurement locations).
	
	\subsection{Damage Estimation}
\begin{figure*}
		\centering
		\subfigure[1992 Big Bear earthquake]{\label{fig:a}\includegraphics[width=75mm]{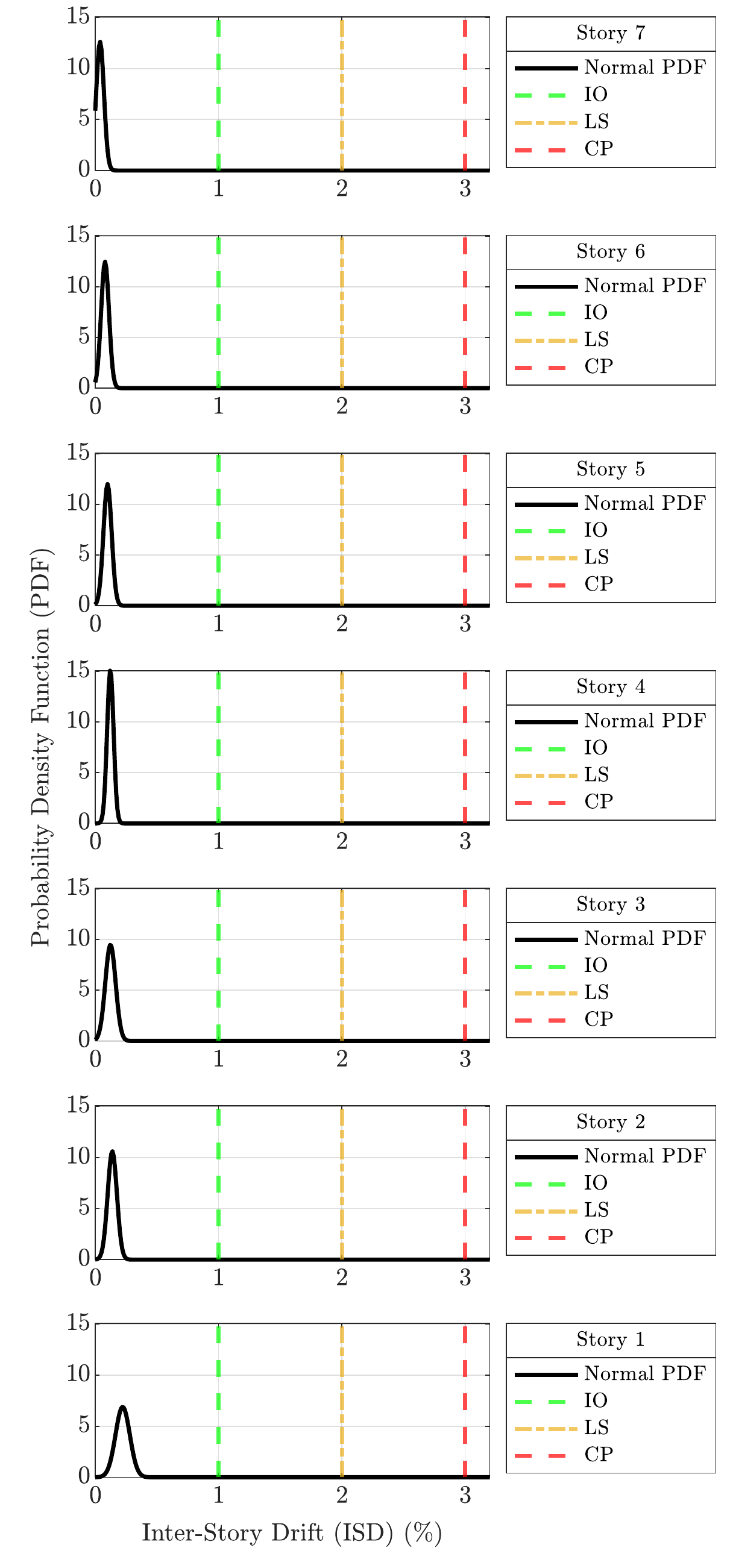}}
		\quad
		\subfigure[1994 Northridge earthquake]{\label{fig:a}\includegraphics[width=75.mm]{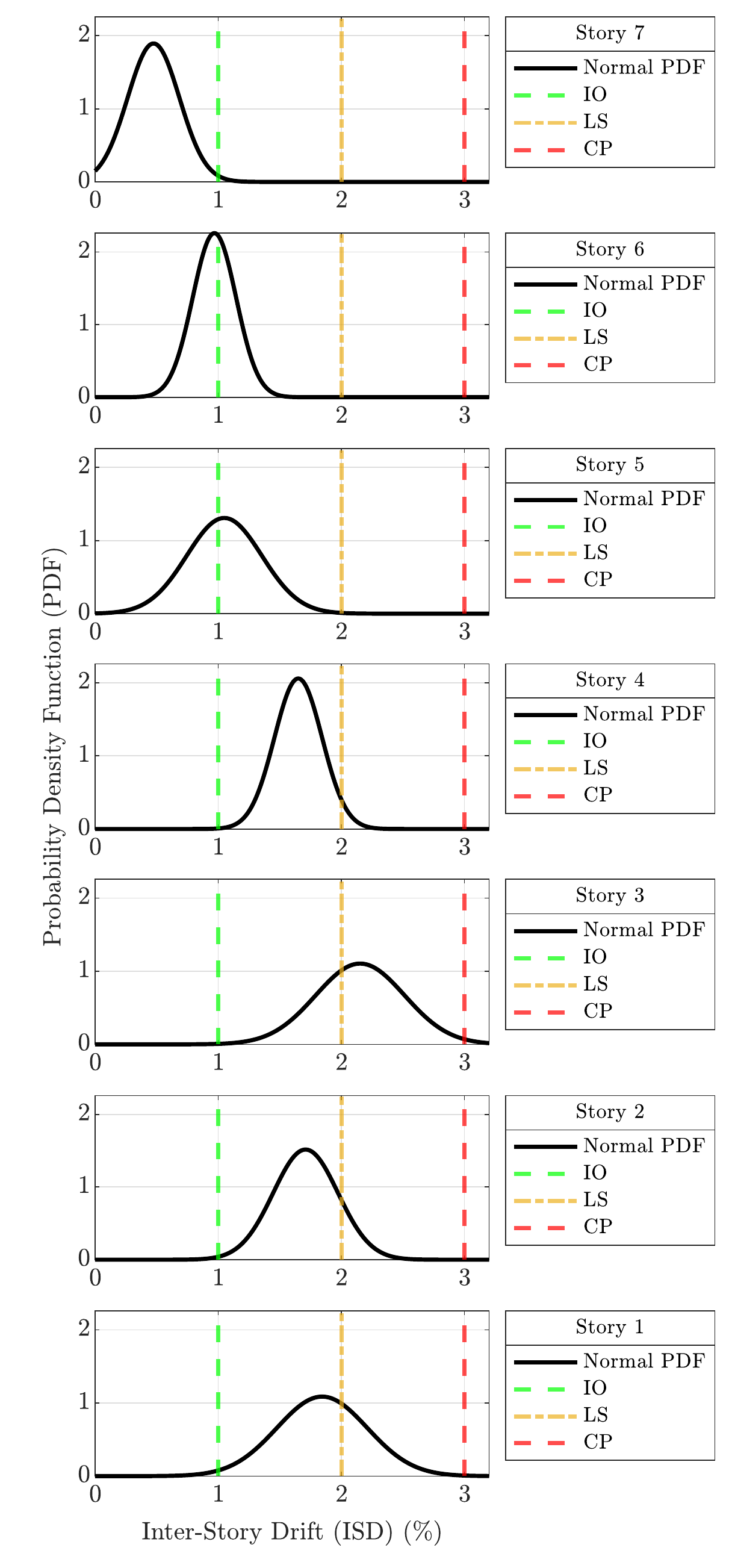}}
		\caption{Reconstructed probability density function for maximum inter-story drift ratios of Van Nuys building during 1992 Big Bear and 1994 Northridge earthquakes}
		\label{fig:PDF}
	\end{figure*}

	Figure \ref{fig:PDF} depicts the estimated PDF of maximum $\text{ISD}$ obtained by fitting normal disrtribution based on first and second moment estimates of maximum $\text{ISD}$ using the OpenSEES-NMBO along with performance-based acceptance criteria of IO, LS, and CP. 
	
\begin{figure*}
		\centering
		\subfigure[1992 Big Bear Earthquake]{\label{fig:a}\includegraphics[width=75mm]{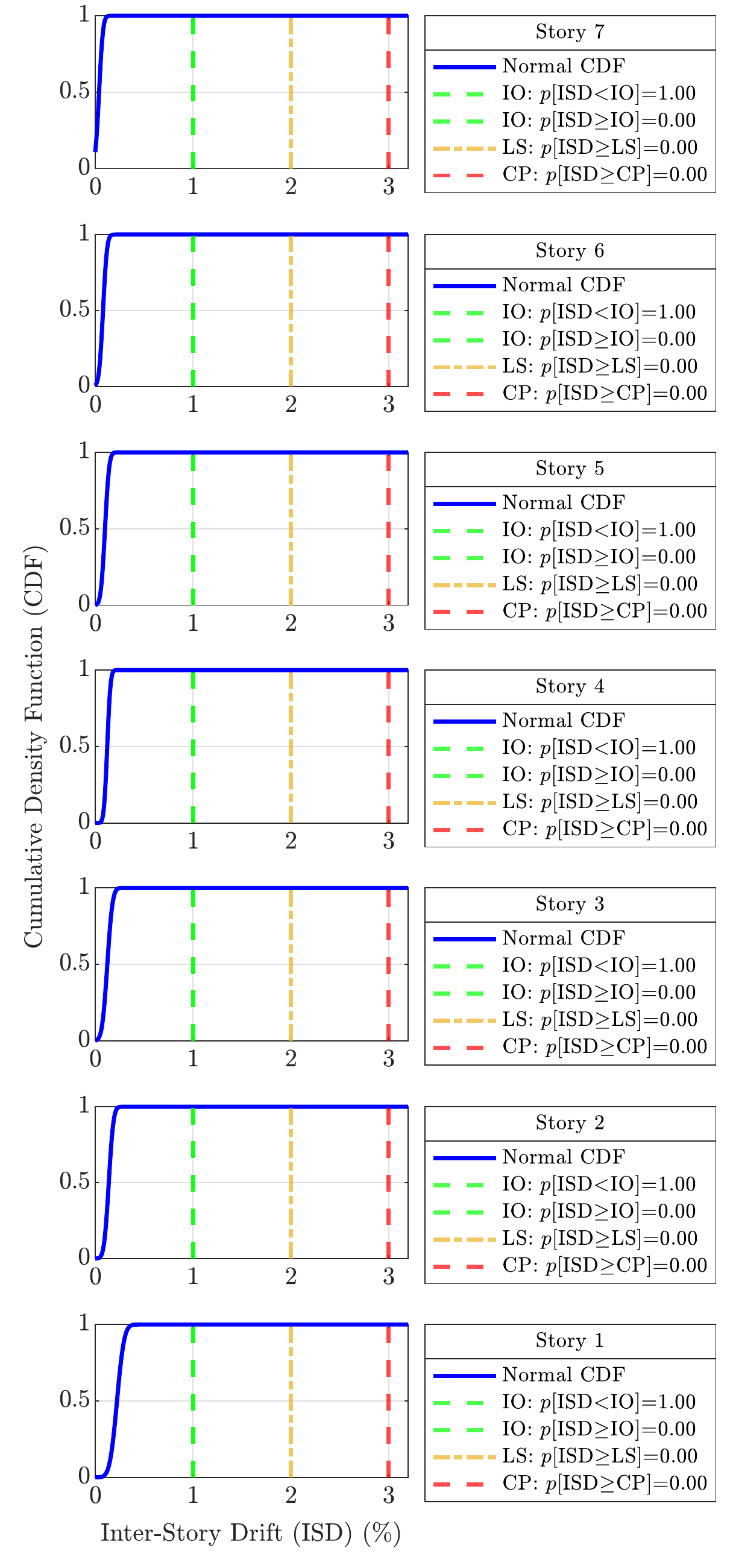}}
		\quad
		\subfigure[1994 Northridge Earthquake]{\label{fig:a}\includegraphics[width=75mm]{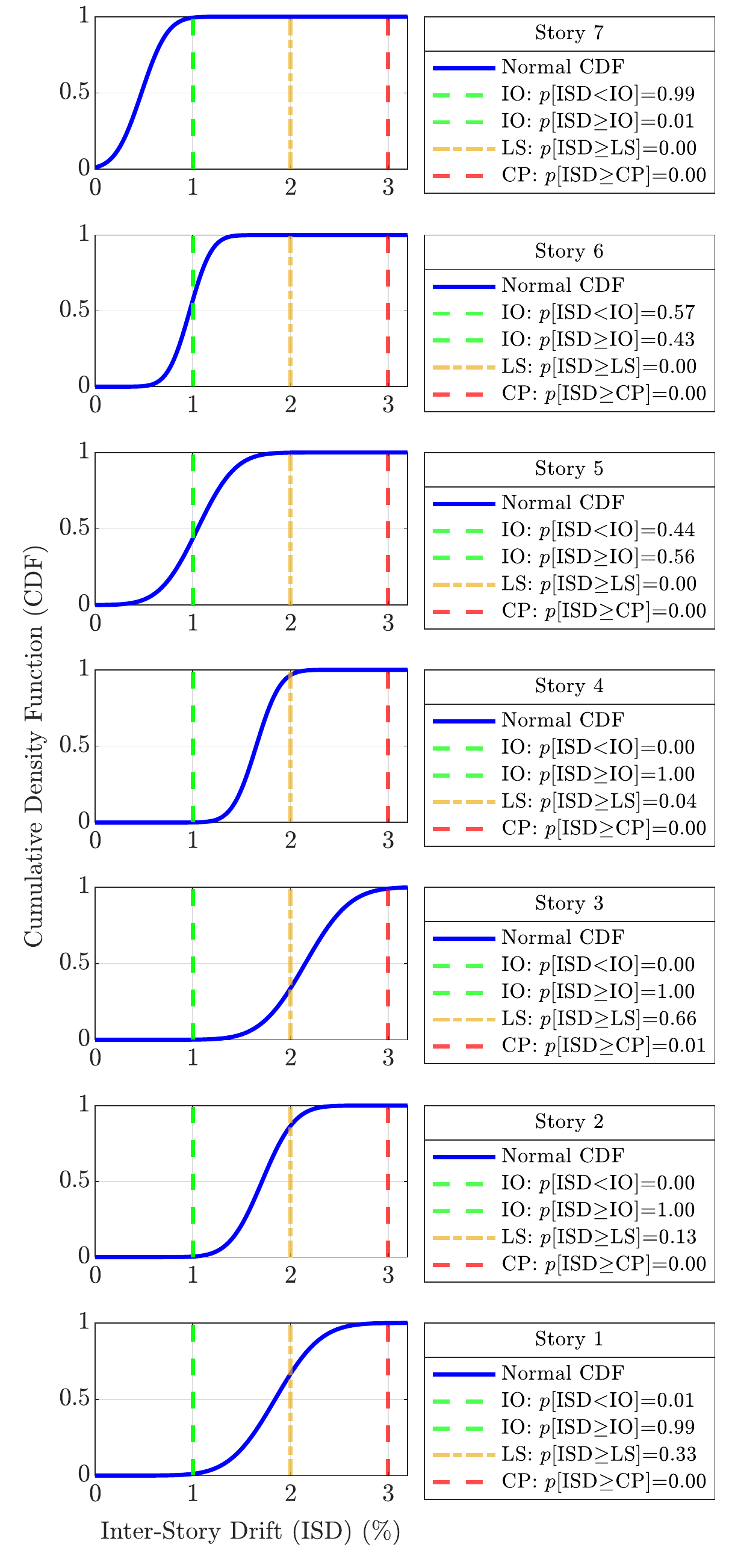}}
		\caption{Reconstructed cumulative density function for maximum inter-story drift ratios and probability of exceeding IO, LS and CP performance levels of Van Nuys building during 1992 Big Bear and 1994 Northridge earthquakes}
		\label{fig:CDF}    
	\end{figure*}

\begin{figure}
		\centering
		\subfigure[1992 Big Bear earthquake]{\includegraphics[width=1\linewidth]{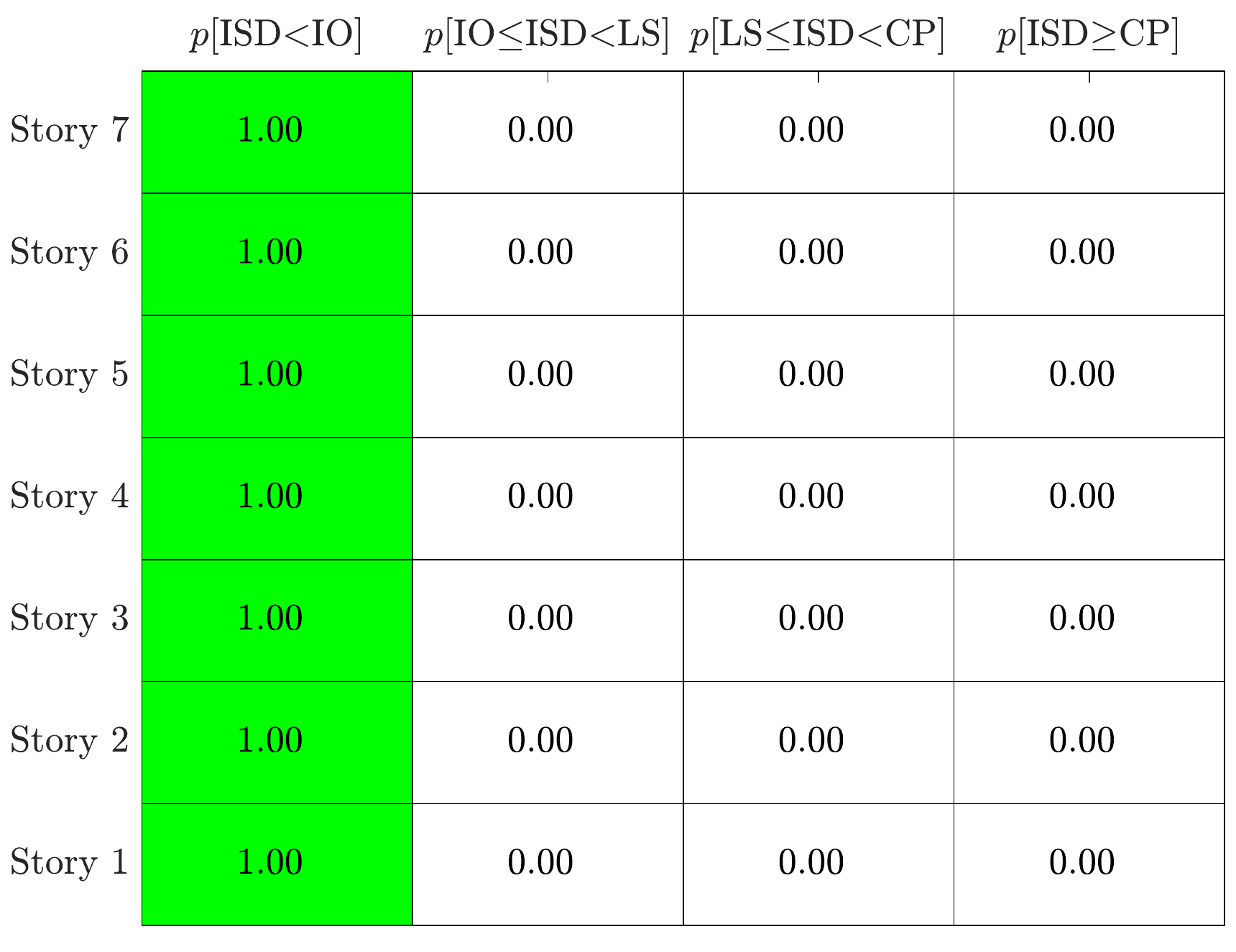}}\\
		
		\subfigure[1994 Northridge earthquake]{\includegraphics[width=1\linewidth]{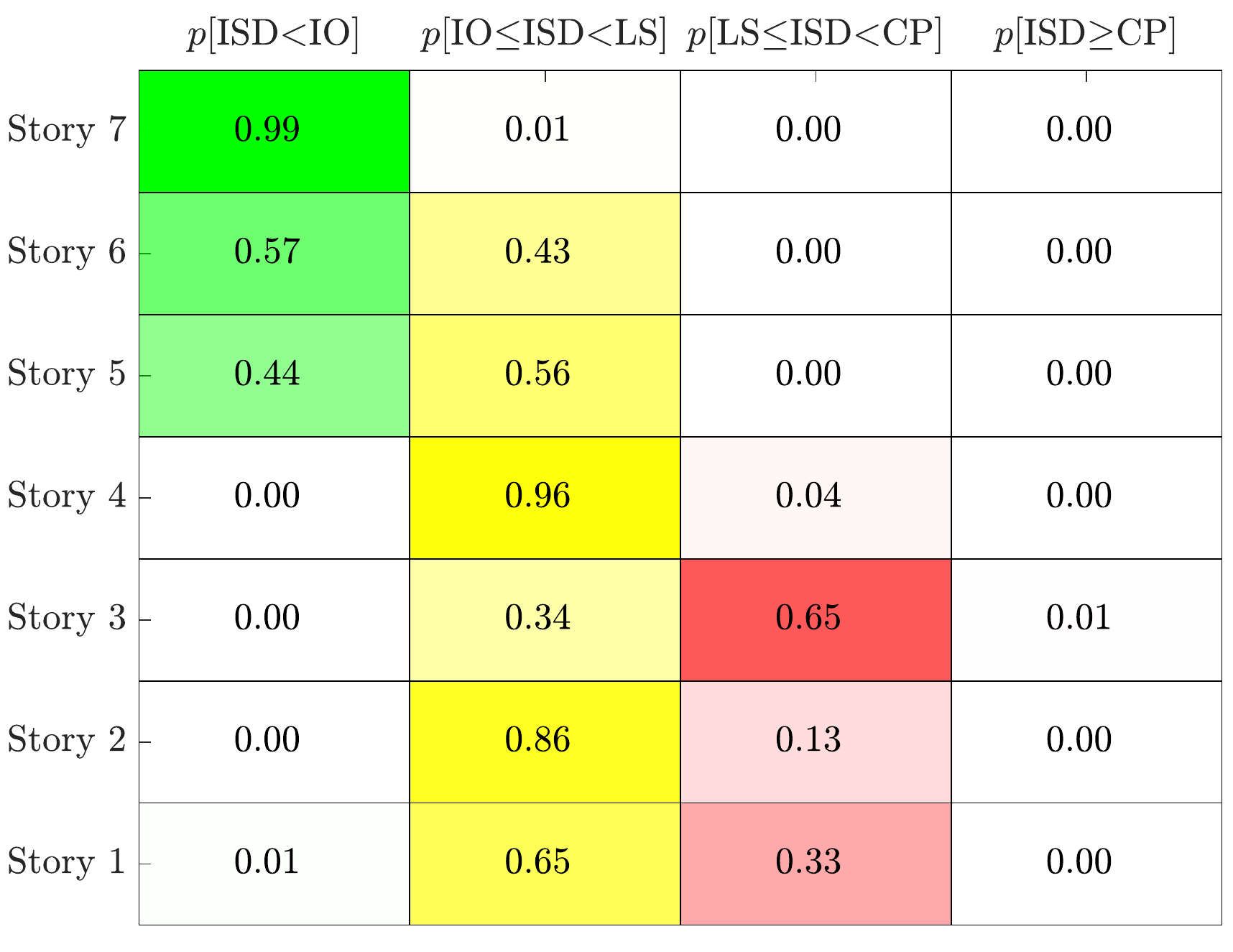}}
		\vspace{-5pt}
		\caption{Story-by-story estimated probability of post-eathquake performance levels of Van Nuys building during 1992 Big Bear and 1994 Northridge earthquakes}
		\label{fig:prob}
	\end{figure}

	Figure \ref{fig:CDF} presents the reconstructed CDF of maximum $\text{ISD}$ for each story during 1992 Big Bear and 1994 Northridge earthquakes. Additionally, performance-based acceptance criteria of IO, LS, and CP, along with the estimated probability of exceeding various performance levels, are presented. Table \ref{tab:DVk} presents the story-by-story probability of exceeding a specific performance level, and Figure \ref{fig:prob} depicts the estimated probability of story-level post-earthquake performance levels of the Van Nuys building.

	{\renewcommand{\arraystretch}{1.3} 
		\begin{table}
			\centering
			\caption{{Story-by-story probability of exceeding specific performance level for the Van Nuys building during 1992 Big Bear and 1994 Northridge earthquakes}}
			\setlength\tabcolsep{3.5pt}
			\scalebox{1}{
				\begin{tabular}{lccccccc}
					\toprule
					\multicolumn{8}{c}{1992 Big Bear earthquake} \\
					\multicolumn{1}{c}{    $k$ (Story)} & 1     & 2     & 3     & 4     & 5     & 6     & 7 \\
					$p[\text{ISD}_k<\text{IO}]$ & 1.00  & 1.00  & 1.00  & 1.00  & 1.00  & 1.00  & 1.00 \\
					$p[\text{ISD}_k\geq\text{IO}]$ & 0.00  & 0.00  & 0.00  & 0.00  & 0.00  & 0.00  & 0.00 \\
					$p[\text{ISD}_k\geq\text{LS}]$ & 0.00  & 0.00  & 0.00  & 0.00  & 0.00  & 0.00  & 0.00 \\
					$p[\text{ISD}_k\geq\text{CP}]$& 0.00  & 0.00  & 0.00  & 0.00  & 0.00  & 0.00  & 0.00 \\
					\hline
					\multicolumn{8}{c}{1994 Northridge earthquake} \\
					\multicolumn{1}{c}{    $k$ (Story)} & 1     & 2     & 3     & 4     & 5     & 6     & 7 \\
					$p[\text{ISD}_k<\text{IO}]$& 0.01  & 0.00  & 0.00  & 0.00  & 0.44  & 0.57  & 0.99 \\
					$p[\text{ISD}_k\geq\text{IO}]$& 0.99  & 1.00  & 1.00  & 1.00  & 0.56  & 0.43  & 0.01 \\
					$p[\text{ISD}_k\geq\text{LS}]$& 0.33  & 0.13  & 0.66  & 0.04  & 0.00  & 0.00  & 0.00 \\
					$p[\text{ISD}_k\geq\text{CP}]$ & 0.00  & 0.00  & 0.01  & 0.00  & 0.00  & 0.00  & 0.00 \\
					\bottomrule
					
				\end{tabular}%
			}
			\label{tab:DVk}%
		\end{table}
		
		\subsection{Post-earthquake Re-occupancy Classification and Decision-Making}
		Table \ref{tab:buildingPL} presents the building level probability of exceeding and classifying specific performance levels for the Van Nuys building, and Figure \ref{fig:VanNuysPL} depicts the estimated probability of post-earthquake building classification for the Van Nuys building.
		\begin{table}
			\centering
			\caption{Building-level Probability of exceeding and classifying specific performance level for the Van Nuys building during 1992 Big Bear and 1994 Northridge earthquakes}
			\scalebox{1}{
				\begin{tabular}{lclc}
					\toprule
					\multicolumn{4}{c}{1992 Big Bear earthquake} \\
					$p[\text{ISD}<\text{IO}]$ & 1.00     & $p[\text{ISD}=\text{IO}]$ & 1.00  \\
					$p[\text{ISD}\geq\text{IO}]$ & 0.00      & $p[\text{ISD}=\text{LS}]$ & 0.00  \\
					$p[\text{ISD}\geq\text{LS}]$ & 0.00      & $p[\text{ISD}=\text{CP}]$ & 0.00  \\
					$p[\text{ISD}\geq\text{CP}]$ & 0.00      & $p[\text{ISD}=\text{C}]$ & 0.00  \\
					\hline
					\multicolumn{4}{c}{1994 Northridge earthquake} \\
					$p[\text{ISD}<\text{IO}]$ & 0.00  & $p[\text{ISD}=\text{IO}]$ & 0.00 \\
					$p[\text{ISD}\geq\text{IO}]$ & 1.00  & $p[\text{ISD}=\text{LS}]$ & 0.19 \\
					$p[\text{ISD}\geq\text{LS}]$ & 0.81  & $p[\text{ISD}=\text{CP}]$ & 0.80 \\
					$p[\text{ISD}\geq\text{CP}]$ & 0.01  & $p[\text{ISD}=\text{C}]$ & 0.01 \\
					\bottomrule
				\end{tabular}%
			}
			\label{tab:buildingPL}%
		\end{table}%
		\begin{figure}
			\centering
			\subfigure[1992 Big Bear earthquake]{\includegraphics[width=1\linewidth]{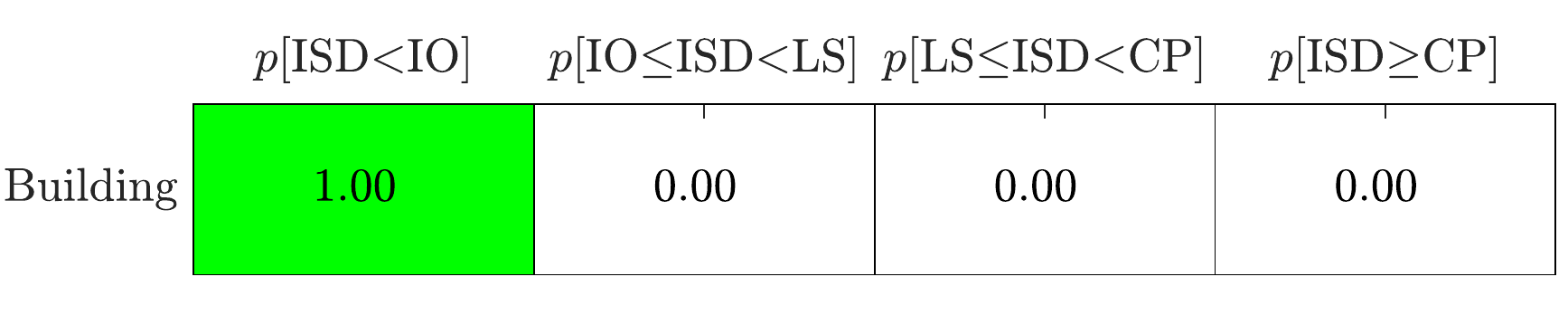}}\\
			
			\subfigure[1994 Northridge earthquake]{\includegraphics[width=1\linewidth]{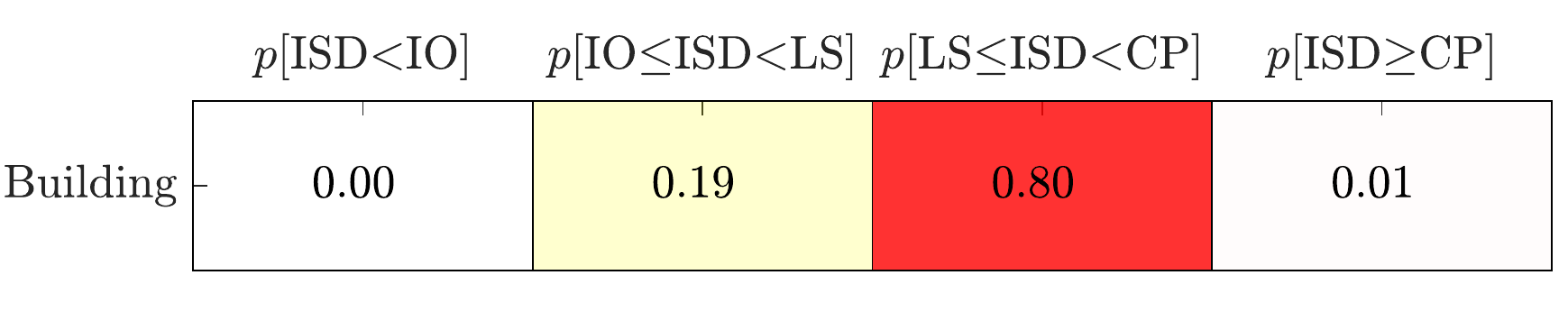}}
			\vspace{-5pt}
			\caption{Building-level estimated probability of post-earthquake performance levels of Van Nuys building during 1992 Big Bear and 1994 Northridge earthquakes}
			\label{fig:VanNuysPL}
		\end{figure}
		\subsection{Discussion on Post-earthquake Assessment and Decision-making Results}
		The preceding sections illustrated that the proposed decision-making framework is capable of combining a refined distributed plasticity FE model and a limited number of response measurements to reconstruct the seismic response of instrumented buildings accurately. Subsequently, the estimated response quantities and their associated uncertainty can be used to reconstruct inter-story drifts with a probabilistic measure of exceeding performance-based acceptance limits and determining the post-earthquake re-occupancy classification of the instrumented building of interest. The $p[\text{DV}_k]$ and $p[\text{DV}]$ estimates during the Big Bear earthquake demonstrate that all the stories remained in the IO performance-level and the building can be classified as IO by a probability of 1. Additionally, the $p[\text{DV}_k]$ estimates during the Northridge earthquake indicate that the higher floors remained IO, and middle and lower floors passed the LS and CP performance levels. In particular, Stories 3 and 4 are classified as LS and CP with the probability of 0.96 and 0.65, which is not consistent with the severe failure of 5 of 9 columns in the fourth story. Also, the $p[\text{DV}_k]$ estimates show that the building can be classified as CP performance-based post-earthquake re-occupancy category by a probability of 0.80. Therefore, these building post-earthquake assessment results are consistent with the building's actual performance and post-earthquake inspection reports following the Big Bear and Northridge earthquakes. In \cite{roohi2019energy}, authors have shown that if the objective is high-resolution story- or element- level damage detection and localization, other damage sensitive response parameters such as element-level dissipated energy, demand-to-capacity ratios, and ductility demand can provide more accurate assessment results compared to inter-story drifts. However, for the application of interest in this paper, the applicability of the proposed framework is validated for rapid performance-based post-earthquake reoccupancy classification and decision-making in the context of a real-world building that experienced severe structural damage during sequential seismic events.
		
		\section{Conclusions}
		This paper develops a performance-based post-earthquake decision-making framework. This framework consists of the following four steps: 1) optimal sensor placement, 2) response reconstruction, 3) damage estimation, and 4) loss analysis. The first step involves the optimal sensor placement of accelerometers. The objective is to select the number and locations of sensors in a way that minimizes the uncertainty in the estimation of displacement response at all stories. The second step is to implement nonlinear model-data fusion and reconstruct probabilistic engineering demand parameters (EDP) in all structural members given the measurements. The third step is to use the estimated EDPs as input to damage models and reconstruct the probability density of damage measures (DM). The DMs are then employed to estimate the probability of decision-variable (DV) exceeding the acceptance criteria from the PBEE concept to determine the post-earthquake re-occupancy category of the instrumented building. Since this concept is developed on a probabilistic basis, the results can also be used to obtain the probability of various losses based on the defined decision variable and loss model. The outcome of this framework can be integrated into a decision-making process by city officials, building owners, and emergency managers.
		
		The framework was successfully implemented using measured data from the seven-story Van Nuys hotel testbed instrumented by CSMIP (Station 24386) during 1992 Big Bear and 1994 Northridge earthquakes. A nonlinear model-based observer of the building was implemented using a distributed plasticity finite element model and measured data to reconstruct seismic response during each earthquake. The estimated seismic response was then used to reconstruct probability and cumulative density of inter-story drifts and determine the performance-based post-earthquake re-occupation category of the building following each earthquake. The performance categories were estimated as IO for Big Bear event and CP for Northridge event.

		\section{Acknowledgement}
		Support for this research provided, in part, by award No. 1453502 from the National Science Foundation is gratefully acknowledged.
		
		\bibliographystyle{unsrt}
		\bibliography{JCSHM-bib}    
		
	\end{document}